\documentclass[11pt]{article}

\usepackage[english]{babel}
\usepackage{amsmath,amsfonts,amssymb,graphicx,color,times,bbm,psfrag,dsfont,
slashed,bigints,bm, verbatim}
\usepackage[dvipsnames]{xcolor}
\def\be{\begin{equation}}
\def\ee{\end{equation}}
\def\bea{\begin{eqnarray}}
\def\eea{\end{eqnarray}}

\def\de{\partial}
\usepackage{xcolor}

\newcommand{\mm}[1]{\textcolor{black}{#1}}
\usepackage{authblk}

\title{Phonon emission by acoustic black holes}

\author[1]{Massimo Mannarelli}
\author[2]{Dario Grasso}
\author[3]{Silvia Trabucco}
\author[4]{Maria Luisa Chiofalo}

\affil[1]{INFN, Laboratori Nazionali del Gran Sasso, Via G. Acitelli,
22, I-67100 Assergi (AQ), Italy}

\affil[2,3,4]{INFN Sezione di Pisa,
Polo Fibonacci, Largo B. Pontecorvo 3, 56127 Pisa, Italy}
\affil[2,3,4]{Dipartimento di Fisica, Universit\`a di Pisa,
Polo Fibonacci, Largo B. Pontecorvo 3}
\date{}
\begin{document}

\maketitle

\begin{abstract}
We present a novel interpretation of the Hawking temperature of acoustic holes, the 
hydrodynamic analogue of standard black holes, by connecting the geometrical properties of the horizon with the distribution function of the spontaneously generated phonons. Using  
covariant kinetic theory to describe the phonon gas emitted by the acoustic hole,  we obtain the correct expression of the Hawking temperature by equating the entropy  loss of the acoustic horizon with the entropy  gain of the phonon gas. In doing this, we assume that the entropy of the acoustic hole is proportional to the area of the horizon, as in standard black holes. Since our method only depends  on the geometrical properties of the acoustic horizon and on the statistical properties of the phonon gas, it is well suited to be extended to standard black holes and to out-of-equilibrium systems.
\end{abstract} 
\section{Introduction}
The transonic flow of a fluid is characterized by the presence of an  acoustic horizon, the hydrodynamic analogue of the event horizon of standard black holes (BHs). In the present paper we report on the investigation performed in~\cite{PhysRevD.103.076001} on how phonons are spontaneously emitted at the horizon of a static acoustic hole and we interpret the mechanism as related to an effective  geometry, determining phonon production and propagation. We perform the computations and discuss the underlying physics within a novel methodology that hinges solely on a kinetic theory approach applied to the fluid analogue of a BH~\cite{PhysRevD.103.076001}~\cite{2008PhRvD..77j3014M}. This conceptually very simple manner to explore the sonic-to-black hole analogy creates deeper understanding of the horizon physics, which can in turn be in principle demonstrated in experiments currently at reach, for example using trapped atomic superfluids as a suited quantum technology platform~\cite{Steinhauer}~\cite{Chin2019}. \\
In an acoustic model of gravity, the emergent metric is fully determined by fluid's quantities such as the density $\rho$, its potential flow $v_\mu$ and the speed of sound $c_s$. For a relativistic flow on a Minkowski spacetime, the analogue metric $g_{\mu\nu}$ can be written as~\cite{2008PhRvD..77j3014M}
\be
g_{\mu\nu}= \eta_{\mu\nu} + (c_s^2-1)v_\mu v_\nu \, ,
\ee
where for simplicity  we neglected an overall conformal factor.
\mm{The inverse metric 
\be 
g^{\mu\nu}=\eta^{\mu\nu} + \left(\frac{1}{c_s^2}-1\right) v^\mu v^\nu \,,
\ee
is computed reminding that $v_\mu$ is a four vector living in flat spacetime, therefore $v^\mu = \eta ^{\mu\nu}v_\nu$.}

Given the acoustic metric, it is possible to study its null geodesics, i.e.~world-lines associated with massless particles. In this acoustic model, such null geodesics  describe on-shell phonon excitations over the background fluid. It is well understood in literature~\cite{Unruh:1980cg} that fluid excitations $\phi$ over a background configuration can be interpreted as quasiparticles following a Klein-Gordon equation
\be 
\frac{1}{\sqrt{-g}}\de_\mu (\sqrt{-g} g^{\mu\nu} \de _\nu \phi )=0\, ,
\ee
where $g=\text{det}(g_{\mu\nu})$. When this massless excitations are originating from the acoustic horizon, they can be depicted as the equivalent of the Hawking radiated particles, predicted in gravity~\cite{Haw1975}. \\
Since we neglect phonon self-interactions, our results are valid sufficiently close to the acoustic horizon. 

\section{Dimensional reduction in a spherical model}
Let us consider a fluid with a smooth convergent flow $v(r)<0$  directed along the $\hat{r}$ direction and speed increasing with $r$: such background configuration provides the acoustic equivalent of the Schwarzschild BH. In spherical coordinates, 
\mm{the background flow is described by the $v_\mu = \gamma(\mm{1}, -v(r),0,0 )$ four vector,}
with $\gamma$ the Lorentz factor. The line element for this configuration reads 
\begin{align} \label{eq:line}
ds^2 = dt^2 (c_s^2-v^2) \gamma^2 + 2 \gamma^2(1-c_s^2) v dr dt - [(1-c_s^2)\gamma^2 v^2 +1]dr^2 - r^2 d \Omega^2\,,
\end{align}
with $d\Omega ^2= d\theta ^2 + \sin ^2 \theta d\varphi^2$ the angular measure. The acoustic horizon of the metric in Eq.~\eqref{eq:line} is given by points at $g_{tt}=0$, i.e. by points at fixed $r$ such that $c_s = v(r_H)$. \\

The dispersion law of a phonon with four momentum $p_\mu =(E,-\mathbf{p})$, can be computed from $g^{\mu\nu}p_\mu p_\nu=0$. 
\mm{Upon decomposing the phonon spatial momentum as a sum of the  parallel and orthogonal components with respect to the background flow, i.e. $\mathbf{p}= \mathbf{p}_r + \mathbf{p}_\perp$, we obtain}
\be
\label{eq:disprel}
E_\pm= \frac{ (\bm{ p} \cdot  \bm{ v}) (1-c_s^2) \pm c_s \gamma^{-2} \sqrt{ (\bm{ p} \cdot  \bm{ v})^2/v^2 + p_\perp^2 \gamma^2(1-c_s^2 v^2)}}{1-c_s^2 v^2}\,.
\ee
We now recall that $\bm v$ is directed along $r$ and the emitted phonons have momenta antiparallel to  the velocity, therefore $\bm{ p} \cdot  \bm{ v} = - p_r v$. Thus
\be
\label{eq:disprel_2}
E_\pm= \frac{ - p_r v (1-c_s^2) \pm c_s \gamma^{-2} \sqrt{ p_r^2 + p_\perp^2 \gamma^2(1-c_s^2 v^2)}}{1-c_s^2 v^2}\,,
\ee
which clarifies that positive energy states correspond to $E_+$. Since phonons are generated by 
horizon fluctuations, they should be emitted  parallel to $\hat{r}$. 
In~\cite{PhysRevD.103.076001} we have argued 

that sufficiently close to the acoustic horizon, phonons of momentum $p^\mu$ can be modeled as quasiparticles with  a factorized distribution function  $f(p^\mu)_\text{ph}= f_\text{\mm{BE}} (p_r)\cdot f_\perp (p_\perp)$, in terms of the Bose-Einstein distribution function and of the Dirac delta distribution function on transverse momentum
\be 
f_\perp (p_\perp) = \left( \frac{2\pi}{L_c} \right) ^2 \delta (p_{\perp}^2)\, ,
\ee
where $L_c$ is some microscopic length scale below which the effective acoustic treatment does not hold. 
The Dirac $\delta$-function determines
an effective dimensional reduction close to the horizon.  In view of a geometrical interpretation of the dimensional reduction mechanism, here we present a different approach.

The direction of phonon propagation corresponds to that of their  group velocity
\be
\bm{v}_g =(v_{gr}\hat{r}, v_{g \perp} \hat{\tau})= \left(\frac{\partial E_+}{\partial p_r} \hat{r}, \frac{\partial E_+}{\partial p_\perp} \hat{\tau}\right)\,,
\ee
where $\hat{\tau}$ is the unit vector of the transverse direction
\mm{and}
\begin{align}
v_{gr} =& \frac{-v (1-c_s^2)+ (1-v^2)c_s p_r/\sqrt{p_r^2+\gamma^2(1-v^2 c_s^2)p_\perp^2} }{1-v^2 c_s^2}\,, \\
v_{g\perp} =& \frac{c_s p_\perp}{\sqrt{p_r^2+\gamma^2(1-v^2 c_s^2)p_\perp^2}}\,,\label{eq:vgperp}
\end{align}
are respectively the radial and transverse components of the group velocity.
The propagation of phonons against the background flow can only happen when $v_{gr} >0$, which occurs when
\be
|p_\perp| < |p_r| \sqrt{\frac{(1-v^2)(c_s^2-v^2)}{v^2(1-c_s^2)}}\,,
\ee
meaning that at the horizon, where $v=c_s$, $p_\perp=0$. \mm{It follows} that at the horizon phonons are emitted radially, indeed for $p_\perp=0$ we have from~\eqref{eq:vgperp} that $v_{g\perp}=0$.
In general,  
the momentum emission angle for any value of $v$ is given by
\be
\alpha=\max\left(2 \arctan\left(\frac{p_\perp}{p_r}\right)\right) = 2 \arctan\left(  \sqrt{\frac{(1-v^2)(c_s^2-v^2)}{v^2(1-c_s^2)}}\right),
\ee
while the corresponding phonon emission angle is \mm{$
\alpha_g=\max\left(2 \arctan\left(\frac{v_{gr}}{v_{g\perp}}\right)\right)
$.}

\section{A thermodynamical argument for an area entropy}
\mm{In~\cite{PhysRevD.103.076001}} we  have  shown  how  for  acoustic holes  the  Hawking  temperature comes \mm{by combining 
covariant kinetic theory 
with geometrical arguments}.  Under the
hypothesis that the black and acoustic hole entropies have an analogous physical origin, we associate to the latter the entropy \be\label{eq:SH} S_H=\frac{A}{4 L_c^2}\,, \ee
in terms of the horizon surface area $A$ and the microscopic length scale $L_c$. We recall that the numerical factor $1/4$ in $S_H$ is computed by matching the expression of $T_H$ with the one found in the non-relativistic limit~\cite{Unruh:1980cg}.
Since the system is isolated, the entropy  variation of the acoustic horizon must equate the entropy variation of the phonon gas, therefore
 \be 
 \label{eq:s}
 dS_H= dS_{\rm ph} \, , 
 \ee
where the phonon entropy variation is given by
\be\label{eq:Sph} 
dS_{\rm ph} =d_g  dV \tilde{s}_{\rm ph} = d_g \tilde{s}_{\rm ph} 4 \pi r_H ^2dr_H \,,
\ee
where $d_g$ is the effective number of phonon degrees of freedom  and $\tilde{s}_{\rm ph}$ is the phonon entropy density, see~\cite{PhysRevD.103.076001} for details. Within this hypothesis, we have determined the expression of the  Hawking temperature:
\be T_H= \frac{1}{2\pi} \left( \frac{c_s-|v|}{1-c_s|v|}\right)'_H\,,
\ee 
which agrees with the standard expression obtained in the non-relativistic limit, see for instance~\cite{Unruh:1980cg,Barcelo:2005fc}.
%by equating the infinitesimal entropy loss of the fluid  with the corresponding phonon entropy gain. 
More in general, the same argument can be retraced in terms of energy balance, \mm{thus}
%To this aim, 
one can write the following thermodynamic relations at equilibrium:
 \be 
 \label{se}
 dS_H= dS_{\rm ph} \, , \qquad  dE_H= dE_{\rm ph} \,,
 \ee
\mm{where $E_H$ is the energy associated to the acoustic horizon and $E_{\rm ph}$ is the phonon gas energy, see~\cite{PhysRevD.103.076001} for more details.} 
These equations can be interpreted as  conservation relations
%, like 
\be dS_H - dS_{\rm ph}=0\, , \qquad dE_H-dE_{\rm ph}=0\,, \ee 
meaning that we are considering a closed system composed by (horizon)+ (fluctuations). We have used the analogy with BH physics to hypothesize the area law for analogue acoustic holes and deduced their properties. We can now read this analogy backwards from analogue to black holes physics and then, Eq.~\eqref{se} returns an interesting perspective. Similarly than with phonons, the spontaneous emission of photons at the event horizon should result in energy and entropy gains of the photon gas equal to the mass and entropy losses of the BH. However, in this case, one complication arises because of the long-range behavior of the gravitational interaction, which may not allow to separate the photon energy density and entropy from those of the BH~\cite{Brout:1995rd}.\\

The dimensional reduction close to the acoustic horizon can be used to justify the assumption in 
Eq.~\eqref{eq:SH},  probably the most subtle point in our approach, which associates an entropy to the acoustic horizon. For a moment, let us put that equation aside: meaning that we are not associating to the horizon any entropy. The effect of the dimensional reduction on thermodynamical functions consists in their temperature dependence as a one dimensional gas rather than a three-dimensional one~\cite{PhysRevD.103.076001}. Moreover, when passing from a three-dimensional to a one-dimensional momentum measure, we are left with a constant cut-off length scale $L_c$. The presence of such a length is significant since it appears in thermodynamical quantities, in particular the covariant phonon entropy density $\tilde{s}_{\rm ph} \propto L_c^{-2}$.

As a consequence of the emission of phonons by the horizon, the phonon entropy density increases by the quantity given in 
Eq.~\eqref{eq:Sph}. Since the system is isolated and we are neglecting phonon self-interactions,  it means that  there is a source of entropy such that
\be \label{eq:Ssource}
dS_\text{source} =d_g  dV \tilde{s}_{\rm ph} = d_g \tilde{s}_{\rm ph} 4 \pi r_H ^2dr_H \,.
\ee
The source entropy variation is thus proportional to the phonon entropy density and the geometrical variation of the horizon. 
Given that $\tilde{s}_{\rm ph} \propto L_c^{-2} $  we also expect that $dS_\text{source} \propto L_c^{-2}$, meaning that $S_\text{source} \propto L_c^{-2}$. However,  the entropy is an adimensional quantity, thus  $S_\text{source}$ has to be  proportional to a ratio between a quantity with dimensions (length)$^{2}$ and $L_c^2$. 
The only (length)$^2$ related with the horizon is the surface area, $A$, of the horizon itself, thus we can write 
\be dS_\text{source} = \kappa \frac{A}{L_c^2}\,,\ee
where $\kappa$ is some dimensionless constant. Fixing $\kappa=1/4$ as in standard BHs we recover Eq.~\eqref{eq:SH}. 
Finally, we note that the dimensional reduction provides a microscopic interpretation of the length $L_c$ for acoustic holes, since it appears in the distribution function as associated with the transverse direction. In gravity, the same role of $L_c$ is played by the 
Planck length \mm{$l_p \propto \sqrt{ G}$}.

In conclusion, our kinetic approach to the physics of acoustic black-holes horizons gains insight on the role of geometry. The Hawking temperature pops out with a conceptually simple meaning of the temperature at which phonons are emitted at the horizon, their energy and entropy changes balancing those of the horizon. This essential idea fosters a different view on standard BHs physics, and can be applied to generate new understanding on dissipative and out-of-equilibrium horizons.

\bibliography{trabucco.bib}

\begin{thebibliography}{1}

\bibitem{PhysRevD.103.076001}
M.~Mannarelli, D.~Grasso, S.~Trabucco, and M.~L. Chiofalo.
\newblock Hawking temperature and phonon emission in acoustic holes.
\newblock {\em Phys. Rev. D}, 103:076001, Apr 2021.

\bibitem{2008PhRvD..77j3014M}
M.~{Mannarelli} and C.~{Manuel}.
\newblock {Transport theory for cold relativistic superfluids from an analogue
  model of gravity}.
\newblock {\em Physical Review D}, 77(10):103014, 2008.

\bibitem{Steinhauer}
J.~Steinhauer.
\newblock {Observation of quantum Hawking radiation and its entanglement in an
  analogue black hole}.
\newblock {\em Nature Physics}, 12:959–965, 2016.

\bibitem{Chin2019}
J.~Hu, L.~Feng, Z.~Zhang, and C.~Chin.
\newblock {Quantum Simulation of Coherent Hawking-Unruh Radiation}.
\newblock {\em Nature Physics}, 15:785–789, 2019.

\bibitem{Unruh:1980cg}
W.G. Unruh.
\newblock {Experimental black hole evaporation}.
\newblock {\em Phys.Rev.Lett.}, 46:1351--1353, 1981.

\bibitem{Haw1975}
S.~W. Hawking.
\newblock Particle creation by black holes.
\newblock {\em Communications in Mathematical Physics}, 43:199--220, 1975.

\bibitem{Barcelo:2005fc}
C.~Barcelo, S.~Liberati, and M.~Visser.
\newblock {Analogue gravity}.
\newblock {\em Living Rev. Rel.}, 8:12, 2005.

\bibitem{Brout:1995rd}
R.~Brout, S.~Massar, R.~Parentani, and Ph. Spindel.
\newblock {A Primer for black hole quantum physics}.
\newblock {\em Phys. Rept.}, 260:329--454, 1995.

\end{thebibliography}
\end{document}